\documentclass[%
reprint,
superscriptaddress,
frontmatterverbose,
 amsmath,amssymb,
prb,
]{revtex4-2}

\usepackage{amsfonts,amsmath,amssymb,amsthm}
\usepackage{dsfont}

\usepackage{graphicx}
\usepackage{dcolumn}
\usepackage{bm}

\usepackage{hyperref}

\hypersetup{
	pdfnewwindow=true,      
	colorlinks=true,       
	linkcolor=blue,          
	citecolor=blue,        
	filecolor=blue,      
	urlcolor=blue,          
}

\usepackage{color}
\usepackage{amsmath}
\renewcommand\boldsymbol{\bm}

\usepackage{lineno}
\makeatother

\newcommand{\be}{\begin{equation}}
\newcommand{\ee}{\end{equation}}

\newcommand{\bk}{{\bm{k}}}

\newcommand{\bR}{{\bm{R}}}

\newcommand{\bx}{{\bm{x}}}

\newcommand{\bp}{{\bm{p}}}

\usepackage{braket}
\usepackage{array}
\usepackage{booktabs}

\newif\ifdraft
\drafttrue
\draftfalse

\begin{document}

\title{Deep Neural Network for Phonon-Assisted Optical Spectra of Semiconductors at finite temperatures}

\author{Qiangqiang Gu}
\email{guqq@ustc.edu.cn}
\affiliation{School of Artificial Intelligence and Data Science, University of Science and Technology of China, Hefei 230026, China}
\affiliation{Suzhou Institute for Advanced Research, University of Science and Technology of China, Suzhou 215123, China}
\affiliation{AI for Science Institute, Beijing 100080, China}
\affiliation{Suzhou Big Data \& AI Research and Engineering Center, Suzhou 215123, China}

\author{Shishir Kumar Pandey}
\affiliation{Birla Institute of Technology \& Science, Pilani-Dubai Campus, Dubai 345055, UAE}

\author{Zhanghao Zhouyin}
\affiliation{Department of Physics, McGill University, Montreal, Quebec, Canada H3A2T8}

\begin{abstract}
{\it Ab initio} based accurate simulation of phonon-assisted optical spectra of semiconductors at finite temperatures remains a formidable challenge, as it requires large supercells for phonon sampling and computationally expensive high-accuracy exchange-correlation (XC) functionals. In this work, we present an efficient approach that combines deep learning tight-binding and potential models to address this challenge with \textit{ab initio} fidelity.  By leveraging molecular dynamics for atomic configuration sampling and deep learning-enabled rapid Hamiltonian evaluation, our approach enables large-scale simulations of temperature-dependent optical properties using advanced XC functionals (HSE, SCAN). Demonstrated on silicon and gallium arsenide across temperature 100-400 K, the method accurately captures phonon-induced bandgap renormalization and indirect/direct absorption processes which are in excellent agreement with experimental findings over five orders of magnitude. This work establishes a pathway for high-throughput investigation of electron-phonon coupled phenomena in complex materials, overcoming traditional computational limitations arising from large supercell used with computationally expensive XC-functionals.
\end{abstract}

\maketitle
\section{Introduction}
The rapid advancement of machine learning (ML) has revolutionized computational materials science, facilitating the development of predictive models across various scales. These models, which range from atomic interaction potentials~\cite{blankNeural1995,behlerGeneralized2007, bartokGaussian2010, thompsonSpectral2015, shapeevMoment2016, zhangDeep2018, schuttSchNet2018, chmielaexact2018, drautzAtomic2019, unkePhysNet2019,mailoafast2019, Gasteiger2020Directional,parkAccurate2021, xieBayesian2021, batznerequivariant2022,musaelianLearning2023} to electronic Hamiltonians in the tight-binding~\cite{guNeural2022, guDeep2024, wangMachine2021, fanObtaining2022, sunMachine2023, mcsloyTBMaLT2023} and Kohn-Sham formalism~\cite{HegdeBowen2017, Schuett2019, nigamEquivariant2022,zhangEquivariant2022, liDeeplearning2022,gongGeneral2023, zhongTransferable2023, zhouyinLearning2024}, have significantly increased the efficiency of material simulations while maintaining \textit{ab initio} level accuracy.
This, in turn, is reflected in machine learning interatomic potentials (MLIPs) showing remarkable success for a wide range of applications, including catalytic processes~\cite{wangStructural2022}, battery optimization~\cite{huImpact2024}, alloy design~\cite{LiuActive2023}, and protein folding~\cite{majewskiMachine2023}, etc. By accurately capturing atomic vibrations, MLIPs have been able to closely replicate experimental spectroscopic data, particularly in Raman and infrared spectra~\cite{sommersRaman2020,xuTensorial2024, zhangDeep2020,gasteggerMachine2017}.
While ML potential models have seen diverse and successful practical applications, the same cannot be said for ML electronic Hamiltonian models. Despite significant advancements in the methodological development of the latter~\cite{guNeural2022, guDeep2024, wangMachine2021, fanObtaining2022, sunMachine2023, mcsloyTBMaLT2023, HegdeBowen2017, Schuett2019, nigamEquivariant2022, zhangEquivariant2022, liDeeplearning2022}, relatively little attention has been given to their practical utility. This is particularly evident in scenarios involving the coupling between electronic and ionic degrees of freedom. There are very few computational studies in this direction~\cite{ep_ml_pccp, cp_ml_prm}, and those that exist are largely limited to zero-point vibrational effects, without accounting for finite-temperature physics.

To bridge this significant gap, we focus on phonon-assisted optical absorption to showcase the potential of ML-based Hamiltonian models in simulating complex electronic processes.  Such an absorption process fundamentally involves the coupling between the electronic and ionic degrees of freedom. It becomes particularly relevant in indirect bandgap materials, where both electron-phonon and electron-photon interactions are required for momentum conservation. These interactions lead to bandgap renormalization, broadening of absorption and emission line shapes, and enable indirect optical transitions.

In this direction, the theory of phonon-assisted indirect optical transitions, developed by Hall, Bardeen, and Blatt~\cite{HallBardeenBlatt1954}, provides foundational insights but does not account for temperature-dependent band structures. This was combined with {\it ab initio} calculations by Noffsinger {\it et al.}~\cite{noffsingerPhonon2012} to incorporate electron-phonon coupling, though temperature-dependent band structure effects still required empirical adjustments.
Phonon-induced renormalization of band gaps and band structures have been studied within density functional theory (DFT)\cite{mariniinitio2008,giustinoElectronphonon2010,giustinoElectronphonon2017}, based on the Allen-Heine theory\cite{allenTheory1976}. M. Zacharias {\it et al.}~\cite{zachariasStochastic2015,zachariasOneshot2016} demonstrated that the quasiclassical Williams-Lax theory~\cite{Williams1951,laxfranck1952} offers a unified framework for calculating optical absorption spectra. This study of the phonon-assisted transitions and band structures renormalization is based on stochastic sampling of phonon normal modes. However, this approach requires large supercells to accommodate normal modes, which can be computationally intensive for \textit{ab initio} calculations, especially for complex systems or when using high-accuracy exchange-correlation (XC) functionals, such as the Heyd-Scuseria-Ernzerhof (HSE) functionals~\cite{heydHybrid2003}. Consequently, previous implementations by M. Zacharias {\it et al.}~\cite{zachariasStochastic2015,zachariasOneshot2016} utilized computationally less demanding GGA functionals with empirical scissor corrections to achieve reasonable accuracy.
This computational constraint limits the applicability of \textit{ab initio} based finite-temperature phonon-assisted optical spectra calculations at large-scale or in high-throughput studies.

The above limitations fall squarely in the realm of ML models, which provide highly efficient simulations. In this work, we address these limitations by presenting a framework that integrates deep potential molecular dynamics (DeePMD)\cite{zhangDeep2018} and deep learning tight-binding (DeePTB)\cite{guDeep2024} models to efficiently and effectively calculate phonon-assisted optical absorption, based on the Williams-Lax theory. By combining DeePMD simulations to capture atomic dynamics with DeePTB for electronic structure and optical properties, our approach allows accurate modeling of temperature-dependent spectral features while significantly reducing computational costs compared to \textit{ab initio} methods.
Taking examples of direct and indirect semiconductors, we
demonstrate the effectiveness of our approach by considering large supercells (up to 4096 atoms) and obtaining the absorption spectra with highly accurate HSE functionals~\cite{heydHybrid2003} and meta-GGA SCAN~\cite{SunSCAN2015} which is in excellent agreement with experimental findings. We emphasize here that obtaining spectra with such a large supercell employing computationally expensive XC-functional was an unachievable task in the past.
This establishes our method's reliability in capturing phonon-assisted processes. Generic nature of our computational approach naturally reproduces the previous finding by M. Zacharias {\it et al.}~\cite{zachariasOneshot2016}  wherein the accurate representation of the ensemble-averaged phonon-assisted absorption spectra can be obtained by individual frames from molecular dynamics simulations using large supercells. 
By establishing a connection between advanced ML models and their practical applications, this approach not only tackles previously intractable challenges in phonon-assisted phenomena but also opens doors to broader applications, including high-throughput design of optoelectronic materials and other temperature-dependent processes in materials science.

\section{Results and Discussion}

\subsection{Methodology and Workflow}

The optical constants of a solid can be derived from the complex dielectric function $\epsilon_1 + i \epsilon_2$. For a system with $N$ atomic coordinate configurations $X=\{\bx_1,\bx_2,\cdots,\bx_N\}$, its $\epsilon_2 (\omega; X)$ can be obtained within the single-particle approximation and dipole approximation as,
\begin{equation}
	\begin{split}
		\epsilon_2(\omega;X) = \frac{2 \pi}{m_\mathrm{e} N_\mathrm{e}} \frac{\omega_{\mathrm{p}}^2}{\omega^2} & \sum_{v, c} \int_{\mathrm{BZ}} \frac{d \bk}{(2 \pi)^3}	\\
		& \times \left|P^X_{c v \bk}\right|^2 \delta\left(\varepsilon^X_{c \bk}-\varepsilon^X_{v \bk}-\hbar \omega\right)
	\end{split}
\label{eq:1}
\end{equation}
where $\omega$ is the photon frequency, $m_\mathrm{e}$ is the electron mass, $N_\mathrm{e}$ is the number of electrons per unit volume and $\omega_{\mathrm{p}} = 4\pi N_\mathrm{e} e^2 / m_\mathrm{e}$ is the plasma frequency, with $e$ being the electron charge. For a given configuration $X$, the crystal momentum $\bk$ and band index $\alpha = v,c$ (valence and conduction bands) label the single-particle electronic states $|\alpha\bk; X \rangle$ with energy $\varepsilon^X_{\alpha\bk}$. The optical matrix element at $clamped$ nuclei $X$ is given by 
$ P^X_{cv\bk} = \langle c\bk; X|\hat{\boldsymbol{n}} \cdot \bp|v\bk; X \rangle$, where $\hat{\boldsymbol{n}}$ is the polarization of the incident light and $\boldsymbol{p}$ is the momentum operator. The real part of the dielectric function $\epsilon_1(\omega,X)$ is obtained using the Kramers-Kronig relation.

\begin{figure}[tbp!]
\includegraphics[width=8 cm]{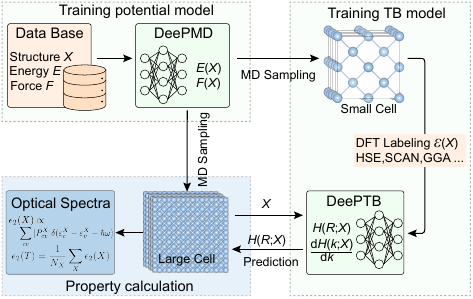}
\caption{Workflow for calculating phonon-assisted optical spectra using DeePMD and DeePTB models. DeePMD generates atomic structures through molecular dynamics simulations. Small-cell structures are then labeled by DFT eigenvalues to train the DeePTB model. Large-cell structures from DeePMD are then used with the trained DeePTB model to predict TB Hamiltonians $H(R)$ and hence the transition matrix for optical spectra calculations.}
\label{fig:workflow}
\end{figure}

According to the Williams-Lax theory~\cite{Williams1951,laxfranck1952}, the imaginary part of the dielectric function at temperature $T$ is evaluated as:
\begin{equation}
\epsilon_2(\omega,T)=\frac{1}{Z}\sum_n e^{-\beta E_n} \left\langle \chi_n| \epsilon_2 (\omega; X) | \chi_n \right\rangle
\label{eq:2}
\end{equation}
where $\beta=1/k_\mathrm{B}T$ with $k_\mathrm{B}$ being the Boltzmann constant. The expectation value is taken at the quantum nuclear state $\chi_n$ with energy $E_n$ under the Born-Oppenheimer approximation. $Z=\sum_n e^{-\beta E_n}$ is the canonical partition function.
Traditionally, Eq.~(\ref{eq:2}) is evaluated through ensemble averaging within the $harmonic$ approximation, where the configurations $X$ are sampled in a large supercell using the normal modes of phonon \cite{zachariasStochastic2015, zachariasOneshot2016,kangFirstprinciples2018}. Although effective in capturing many aspects of phonon-assisted optical absorption, this approach may not fully account for anharmonic effects, particularly at higher temperatures, which may result in strong renormalization of phonon spectra, in turn, resulting in drastic moderation of electronic properties as mentioned earlier~\cite{an_sc1, an_sc2, an_fe,an_te}.

In this study, as illustrated in Fig.~\ref{fig:workflow}, we employ molecular dynamics (MD) simulations to directly sample atomic displacements at finite temperatures. This method offers a more comprehensive sampling of configuration space and may naturally incorporating anharmonic effects, and is expressed as:
\begin{equation}
\epsilon_2(\omega,T)= \frac{1}{N_X} \sum_{X \in X(t)} \epsilon_2 (\omega; X)
\label{eq:3}
\end{equation}
where $X(t)$ represents the configurations sampled from the MD trajectory, and $N_X$ is the number of sampled configurations.

Figure \ref{fig:workflow} illustrates our methodology, which leverages deep learning models to enhance computational efficiency for phonon-assisted optical absorption calculations. We employ a DeePMD model~\cite{zhangDeep2018}, trained on a dataset of atomic configurations $X$ along with their corresponding energies $E(X)$ and forces $F(X)$. This model is used to generate atomic structures through MD simulations at finite temperatures, serving dual purpose: generating small-cell structural data, annotated with DFT-calculated electronic eigenvalues $\varepsilon^X$ for training the DeePTB model~\cite{guDeep2024}, and producing large-cell structural data for comprehensive configuration space sampling for the evaluation of Eq. (\ref{eq:3}).  The trained DeePTB model enables efficient and accurate prediction of the TB Hamiltonian $H(\mathbf{R}; X)$ for every sampled configuration $X$ in MD simulations, where $\mathbf{R}$ represents the lattice vector. This accurate and rapid access to $H(\mathbf{R}; X)$ facilitates the investigation of correlations between electronic and ionic degrees of freedom, which is crucial for systems or phenomena involving strong electron-phonon coupling~\cite{zhangDeep2018, guDeep2024}. Specifically, for phonon-assisted optical transitions, the transition matrix for a configuration with clamped nuclei $X$ can be obtained using the DeePTB-predicted Hamiltonian:
\begin{equation}
\boldsymbol{P}^X_{cv\bk}(\omega,T)= \langle c\bk; X | \frac{m_\mathrm{e} \partial H(\bk;X)}{\hbar\partial \bk} |v\bk; X \rangle\label{eq:4}
\end{equation}
where $H(\bk;X)$ can be obtained from Fourier transformation of the $H(\bR;X)$. By combining DeePMD and DeePTB models, we can evaluate Eqs.~(1), (3), and (4) for optical spectra calculations.

\subsection{Optical Absorption Calculations}
Following the introduction of our methodology combining DeePMD and DeePTB, we demonstrate its effectiveness by applying it to phonon-assisted optical absorption in representative semiconductors with indirect (Si) and direct (GaAs) bandgaps. Our approach excels at accurately capturing complex phonon-assisted processes while maintaining computational efficiency. The details about the data preparation and model training are shown in the supplementary materials (SM). The following sections will detail our findings, highlighting the method’s accuracy, insights into physical phenomena, and potential applications in materials science.

\begin{figure}[t!]
\includegraphics[width=8.0 cm]{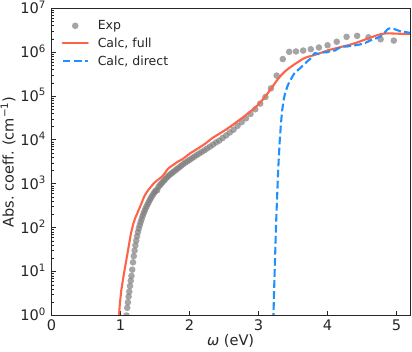}
\caption{Absorption coefficient of silicon (Si) at 300 K. Blue dashed lines represent calculations with atoms clamped at their equilibrium positions. Red solid lines denote calculations with atom configurations sampled in MD simulations at 300K. Experimental data~\cite{greenOptical1995} for Si are shown as grey discs. Calculations were performed using $8 \times 8 \times 8$ conventional supercells and $2\times 2 \times 2$ $\bk$-grid for Brillouin zone (BZ) sampling with a Gaussian broadening of 30 meV.}
\label{fig:Si_300K}
\end{figure}

\begin{figure*}[t]
\centering
    \includegraphics[width=14.0 cm]{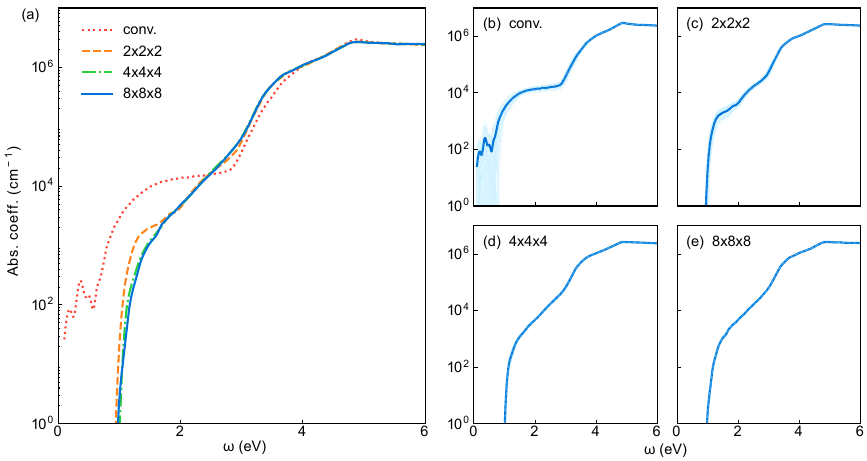}
    \caption{Convergence of supercell size and number of snapshots (a) The optical absorption spectra of Si were calculated from ensemble averages with different supercell sizes. Calculated ensemble averaged spectra considering five snapshots for (b) conventional, (c) 2 $\times$ 2 $\times$ 2, (d) 4 $\times$ 4 $\times$ 4, and (e) 8 $\times$ 8 $\times$ 8 cell size. The shaded region represents the deviation of spectra from the ensemble-averaged one. As the cell size increases, the deviation in spectra gets substantially reduced.}
    \label{fig:Si_diff_size}
\end{figure*}

Figure \ref{fig:Si_300K} compares the calculated optical absorption coefficient of Si at 300K with experimental data~\cite{greenOptical1995}. The absorption coefficient was computed as $\kappa(\omega,T)=\omega \epsilon_2(\omega;T)/cn(\omega;T)$, where $c$ is the speed of light and $n(\omega;T)={\frac{1}{2}[\epsilon_1(\omega;T)+\sqrt{\epsilon_1^2(\omega;T)+\epsilon_2^2(\omega;T)}]}^{1/2}$ is the refractive index. To fully capture atomic dynamics effects on optical absorption due to electron-phonon coupling, we used the DeePMD model to run MD simulations on an  $8\times 8\times 8$ supercell (4096 atoms) and computed absorption spectra for multiple snapshots using the DeePTB model, trained on HSE-functional~\cite{heydHybrid2003} DFT eigenvalues for accurate Si electronic structure. 
For comparison, we also present absorption coefficients evaluated using the DeePTB Hamiltonian with atoms clamped at equilibrium positions (blue dashed line) in the same $8\times 8\times 8$ conventional supercell.
The spectrum calculated at equilibrium positions exhibits an onset at the direct gap $\sim$ 3.3 eV, corresponding to the direct $\Gamma_{25^\prime} \to \Gamma_{15}$ transition in the primitive unit cell band structure. Notably, the indirect absorption between 1.1-3.3 eV observed experimentally~\cite{greenOptical1995} is absent in this calculation. In contrast, the optical absorption coefficient evaluated using the ensemble average defined by Eq.~(\ref{eq:3}) correctly reproduces absorption below the direct gap, showing excellent agreement with experimental data across a wide range of photon energies and over five orders of magnitude. The underestimation of the $E_1$ transition strength around 3.3 eV in our calculations, is attributed to the omission of excitonic effects (electron-hole interactions), which are known to increase the oscillator strength of the $E_1$ peak~\cite{benedictTheory1998}.

Furthermore, our convergence study on supercell size reveals crucial insights into the nature of phonon-assisted absorption calculations. 
As shown in Fig.~\ref{fig:Si_diff_size}(a), the absorption spectra converge progressively as the supercell size increases from the conventional unit cell to $L \times L \times L$ ($L=2,4,8$) multiples. The results for $L=4$ and $8$ are indistinguishable, indicating that a supercell size with $L \geq 4$ is sufficient for accurate representation of phonon-assisted absorption for Si.
To demonstrate the fact that a single configuration can statistically represent the ensemble average of the phonon-assisted absorption process, in Fig.~\ref{fig:Si_diff_size} (b), (c) and (d), we performed ensemble averaging using five snapshots for each supercell size. One can observe from these figures that with increase of the cell size, a significant reduction in the deviation among absorption spectra (shaded region) from different MD configuration snapshots occurs. For $L = 4$ and $8$ in Fig.~\ref{fig:Si_diff_size} (d) and (e), this deviation becomes negligible implying that for sufficiently large supercells, a single configuration can statistically represent the ensemble average of the phonon-assisted absorption process. This finding is consistent with previous studies~\cite{zachariasOneshot2016}, where the one-shot calculation is proofed through carefully designed combinations of specific phonon modes. However, we would like to emphasize here that our approach is quite generic in nature as single MD snapshots that can be used to calculate the spectra do not have a specific atomic displacement pattern, thus eliminating the need for any designing of atomic configurations. Hence, our approach can offer potential computational strategies for studying more complex materials with reduced computational cost.
 
\begin{figure}[t!]
    \includegraphics[width=8 cm]{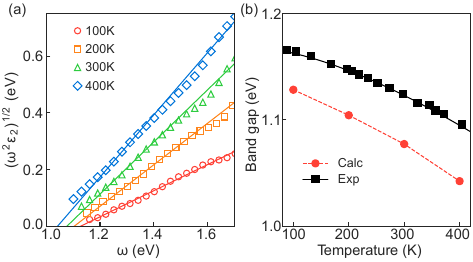}
    \caption{Temperature-renormalized indirect band gaps of Si. (a) Tauc plot for determining the indirect band gap as a function of temperature. The markers, in different colors, represent $(\omega^2\epsilon_2)^{1/2}$ at various temperatures, with solid lines of the same color showing the corresponding linear fits. The indirect band gap is obtained from the intercept with the horizontal axis. (b) The temperature dependence of the indirect band gap from the present theory (red discs) and the experimental data~\cite{alexTemperature1996} (black squares). The solid and dashed lines serve as a guide to the eye.}
    \label{fig:Si_ind_bandgaps}
\end{figure}

To deepen our understanding of Si's temperature-dependent electronic properties, we extended our study to cover a range from 100K to 400K, focusing on the evolution of the indirect band gap—a key factor influencing semiconductor behavior and device performance.
For indirect band gap semiconductors like Si, according to the Williams-Lax theory, the imaginary part of the temperature-dependent dielectric function, $\epsilon_2(\omega; T)$, shows an absorption onset at the temperature-dependent band gap $\Delta_T$. Moreover, based on the standard parabolic approximation for band edges in three-dimensional solids~\cite{bassani1975electronic}, $\epsilon_2(\omega; T)$ near the absorption onset satisfies the relation:
$
[\omega^2 \epsilon_2 (\omega; T)]^{1/2} \propto (\hbar\omega - \Delta_T)
$, which supports the Tauc plots~\cite{taucOptical1966}, a widely used experimental method for band gap determination.
Figure \ref{fig:Si_ind_bandgaps}(a) presents Tauc plots for Si across the 100-400K range. As expected, linear behavior is observed over an energy range spanning nearly 1 eV from the absorption onset. The indirect band gaps were extracted from the intercepts of these linear fits for each temperature, as shown in Fig. \ref{fig:Si_ind_bandgaps}(b) (red discs), along with experimental data (black squares).
The calculated band gaps align well with experimental trends, accurately capturing the temperature-induced narrowing. However, a consistent offset between our calculations and experimental values is observed, likely due to the inherent limitations of the DFT calculations used for training our DeePTB models. Notably, our DeePTB model was trained on band structure data from computationally intensive HSE hybrid functional calculations on small cells, which, in turn, enables efficient and accurate calculations for large supercells due to its predictive capability.

\begin{figure}[t!]
    \includegraphics[width=8.0 cm]{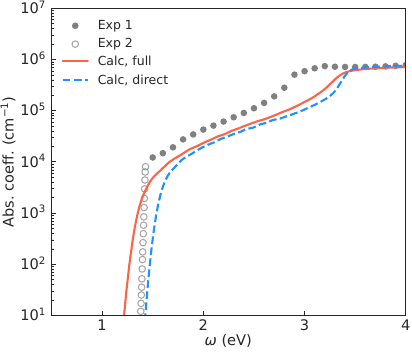}
    \caption{Absorption coefficient of GaAs at 300 K. Blue dashed lines represent calculations with atoms clamped at their equilibrium positions. Red solid lines denote calculations with atom configurations sampled in MD simulations at 300K. Experimental data for GaAs are shown as grey~\cite{sturgeOptical1962} and black~\cite{aspnesDielectric1983} discs. Calculations were performed using $8 \times 8 \times 8$ conventional supercells and $5\times 5 \times 5$ $\bk$-grid for Brillouin zone (BZ) sampling  with a Gaussian broadening of 50 meV.}
    \label{fig:GaAs_300K}
\end{figure}

To further demonstrate the versatility of our method, we applied it to GaAs, a representative direct band gap semiconductor widely used in optoelectronics. MD simulations were conducted at 300K using the DeePMD model on an $8 \times 8 \times 8$ supercell to sample atomic configurations. The DeePTB model, trained on SCAN functional eigenvalues of conventional unit cell configurations, was adjusted with a 1.0 eV scissor correction to align with experimental band gaps.
In Fig.~\ref{fig:GaAs_300K}, we show a comparison of our calculated absorption spectra of GaAs at 300K (red solid line)  with that of experiments~\cite{sturgeOptical1962, aspnesDielectric1983}. For quantifiable comparisons, absorption coefficients calculated for atoms clamped at their equilibrium positions are also shown (blue dashed line). The results reveal that although GaAs is a direct band gap material, phonon-assisted processes notably impact its optical properties, particularly in the region below the fundamental band gap. While the absorption spectrum from MD-sampled and equilibrium positions remains consistent in shape—typical for a direct band gap material, the phonon-assisted spectrum shows a redshift, consistent with the previous finding~\cite{zachariasOneshot2016}.
Our results are in good agreement with experimental data over a broad energy range, accurately capturing the absorption onset at the direct gap and subtle high-energy features. However, our calculated absorption coefficient slightly underestimates experimental values, likely due to the exclusion of excitonic effects and limitations in the DFT-based description of effective masses. Nonetheless, our method effectively reproduces the phonon-assisted absorption below the direct gap and accurately predicts the effects of electron-phonon interactions, a crucial feature in real semiconductors.

\subsection{Summary and Discussion}
In summary, we have presented a novel and efficient method for calculating phonon-assisted optical absorption spectra in semiconductors by combining the deep potential molecular dynamics and deep learning tight-binding models employing DeePMD and DeePTB packages. Our approach accurately reproduces temperature-dependent optical properties of indirect (Si) and direct (GaAs) gap semiconductors which are in excellent agreement with experimental data across a wide range of temperature (100-400 K). Notably, we find that for large supercells, individual frames from MD simulations can accurately represent the ensemble average of phonon-assisted absorption spectra.
This insight, coupled with our method's ability to leverage advanced functionals like HSE and SCAN, makes possible calculations on large supercells, which was previously inaccessible. Such a capability pushes the boundaries of computational feasibility in studying phonon-assisted optical processes. Furthermore, the efficiency and accuracy of our approach open new avenues for the high-throughput computational design of optoelectronic materials. As a future prospect, our method can be extended to calculate the absorption spectra of more complex systems such as alloys and nanostructures, as well as to other phonon-assisted phenomena. By bridging the gap between advanced machine learning models and practical applications, we provide a powerful tool for addressing complex materials science problems previously considered computationally intractable.


\section*{Code availability}
DeePTB package is publicly available as an open-source package from GitHub under \url{https://github.com/deepmodeling/DeePTB}.



%

\end{document}